\documentclass[%
 aip,
 amsmath,amssymb,
 reprint,%
]{revtex4-1}

\usepackage{graphicx}
\usepackage{dcolumn}
\usepackage{bm}

\usepackage[utf8]{inputenc}
\usepackage[T1]{fontenc}
\usepackage{mathptmx}
\usepackage{etoolbox}
\usepackage[version=4]{mhchem}
\usepackage{stfloats}
\usepackage{graphicx}
\usepackage{subcaption}
\usepackage{float}
\usepackage{placeins}
\usepackage[justification=centering,singlelinecheck=false]{caption}

\makeatletter
\def\@email#1#2{%
 \endgroup
 \patchcmd{\titleblock@produce}
  {\frontmatter@RRAPformat}
  {\frontmatter@RRAPformat{\produce@RRAP{*#1\href{mailto:#2}{#2}}}\frontmatter@RRAPformat}
  {}{}
}%
\makeatother
\begin{document}

\preprint{AIP/123-QED}

\title{Time-dependent density functional theory investigation of the formation of \ce{H3+} from alkanes}
\author{C. Jiang}
\affiliation{Department of Physics and Astronomy, Vanderbilt
University, Nashville,
Tennessee 37235, United States}
\author{Samuel S. Taylor}
\affiliation{Department of Physics and Astronomy, Vanderbilt
University, Nashville,
Tennessee 37235, United States}
\author{Kedong Wang}
\affiliation {\it School of Physics, Henan Normal University, Xinxiang
453007, People’s Republic of China}
\author{Cody L. Covington}
\affiliation{ Department of Chemistry, Austin Peay State University,
Clarksville,
Tennessee 37044, United States}
\author{Kalman Varga}
\email{kalman.varga@vanderbilt.edu (corresponding author)}
\affiliation{Department of Physics and Astronomy, Vanderbilt
University, Nashville,
Tennessee 37235, United States}

\date{\today}

\begin{abstract}
The formation of \ce{H3+} from ethane, propane, and butane dications was investigated with time-dependent density-functional theory (TDDFT) simulations.  
This approach offers the benefit of simultaneously addressing nuclear
and electronic dynamics, enabling the investigation of electronic
excitation, charge transfer, ionization, and nuclear motion.
For each dication we determined the ground-state HOMO, the branching ratios of all dissociation channels, and the mechanism leading to \ce{H3+}. The simulated branching ratios for ethane and propane are similar, while butane is markedly lower. Ethane follows the minimum-energy pathway (MEP) proposed previously; propane forms \ce{H3+} mainly via \ce{H2} roaming.  
In butane, \ce{H3+} appears only through the MEP within the present trajectory set; roaming \ce{H2} was not productive under the same conditions.  

\end{abstract}

\maketitle

\section{\label{sec:level1}Introduction}

H$_3^+$, a triangular three‐proton ring, is a cornerstone of interstellar chemistry. By protonating molecules such as \ce{H2O}, H$_3^+$ triggers ion–molecule reactions that create more complex species and ultimately promote star‐formation processes.\cite{ref1} In astrophysical environments, H$_3^+$ is believed to form primarily through cosmic‐ray ionization of molecular hydrogen,\cite{ref2}
\[\ce{H2+ + H2 -> H3+ + H}\],
yet in the interstellar medium the observed column densities of H$_3^+$ exceed classical gas‐phase predictions by orders of magnitude.\cite{ref3,ref4,ref5,ref6,ref7} The previously ignored coulomb‐driven fragmentation of doubly ionized organic molecules accounts for only a minor contribution to the extra \ce{H3+} abundance. \cite{ref17}However, this unimolecular reaction is particularly interesting because it involves the sequential cleavage and formation of at least three bonds on an ultrafast timescale, which is essential for our understanding of molecular physics and photochemistry. Studying the formation of \ce{H3+} from organic molecules also provides crucial insight into the dynamics and mechanisms of \ce{H3+} reactivity in astrophysical environments, effectively representing the reverse “half-collision” of its protonation of molecules, atoms, and ions.\cite{ref21}

A broad range of molecules - including alkanes, alkenes, halogenated and pseudohalogenated compounds, alcohols 
(and their isotopologues), thiol analogues (\ce{OH-SH}), cyclic
hydrocarbons, and amino‐containing molecules 
has been shown to yield H$_3^+$ after double ionization 
\cite{ref8,ref9,ref10,ref11,ref12,ref13,ref14,ref15,ref15,ref16,ref17,ref18,ref19,ref20,ref21,ref22,ref23,ref24,ref25,ref26,ref27,ref28,ref29,ref30,ref31,ref32,ref33,ref34,
zhou_ultrafast_2023,mishra_direct_2024,kwon_what_2023,ando_coherent_2018,wang_role_2021,zhang_formation_2019}.
In a  pioneering work, Kraus \textit{et al.}\cite{ref9} combined femtosecond‐laser ionization with high‐level quantum‐chemical calculations to elucidate H$_3^+$ formation from the \ce{C2H6^2+} dication. They proposed a minimum‐energy‐path (MEP) mechanism: an initial hydrogen migration generates a \ce{[CH2-CH4]^2+} intermediate, from which H$_3^+$ is released via a transition state resembling a neutral \ce{H2} moiety bound to an ethylene dication. Subsequent 300 eV electron‐impact studies by Zhang \textit{et al.}\cite{ref11} confirmed the MEP and, crucially, revealed an additional \ce{H2}–roaming channel in which the neutral \ce{H2} wanders for $\sim$100 fs before abstracting a third proton. This \ce{H2}–roaming pathway has since been identified across a wide range of organic molecules—from alcohols to cyclic hydrocarbons—underscoring its general importance in double‐ionization chemistry \cite{ref13,ref16,ref18,ref19,ref20,ref21,ref24,ref25,ref27,ref29,ref33,ref34}. Extending these insights, Ekanayake \textit{et al.}\cite{ref21} used femtosecond‐laser ionization and ab initio molecular dynamics (AIMD) simulations on a series of alcohols, showing that H$_3^+$ formation yield decreases with chain length. More recently, Kwon \textit{et al.}\cite{ref16} investigated propene and cyclopropane with strong‐field ionization and AIMD, proposing for cyclopropane a sequence of ring opening, hydrogen migration, \ce{H2} roaming, and final proton abstraction. Efforts have also been made to control the formation of H$_3^+$ by shaping the laser pulse used to produce double ionization \cite{ref10,ref12}.

However, H$_3^+$ formation from doubly ionized longer‐chain alkanes—specifically propane and butane—remains unexplored. We therefore ask whether the established MEP and \ce{H2}-roaming pathways persist in these larger molecules and whether the declining‐yield trend seen for alcohols holds for pure alkanes.

In this work, we report time‐dependent density‐functional theory \cite{PhysRevLett.52.997,runge_density-functional_1984,ullrich_time-dependent_2012} 
simulations on ethane, propane, and butane. We analyze the reaction pathways, formation time scales, and branching ratios of H$_3^+$ generated from ethane, propane, and butane, and benchmark these results against the existing experimental data and theoretical studies on the double ionization of organic molecules. To the best of our knowledge, this is the first TDDFT study of H$_3^+$ formation from organic molecules. Previous AIMD simulations typically used Born–Oppenheimer molecular dynamics (BOMD) \cite{ref16,ref18,ref19,ref20,ref34}. BOMD moves nuclei on a single ground‐state surface, re‐solving the electronic Schrödinger equation with a self‐consistent‐field (SCF) cycle at every step \cite{ref35}. This adiabatic scheme eliminates non‐adiabatic electron–nuclear feedback and quickly becomes expensive for large trajectory sets \cite{ref36}. In contrast, real‐time TDDFT propagates electrons and nuclei together under a time‐dependent Kohn–Sham Hamiltonian, capturing charge migration, state mixing, and energy flow on attosecond–femtosecond scales without pre‐built surfaces or repeated SCF loops \cite{ref37,ref38}.
This method has already delivered accurate ultrafast dynamics across diverse systems.\cite{ref39,ref40,ref41,ref42,ref43,ref44,ref45} 
Accordingly, we adopt real‐time TDDFT for the present investigation.

Our recent research
\cite{taylor2024timedependentdensityfunctionalstudyhydrogen}
investigated hydrogen adsorption and scattering
behavior on graphene surfaces using two distinct computational
methods: adiabatic time-dependent density functional theory (TDDFT)
and non-adiabatic ab initio molecular dynamics. The comparative
analysis demonstrated substantial discrepancies in both the predicted
nuclear motion pathways and the electronic excitation dynamics of
hydrogen atoms. These findings underscore the critical importance of
incorporating non-adiabatic effects when simulating such surface
interactions. Specifically, adiabatic methodologies, which operate
under the assumption that electrons continuously occupy their ground
state, fail to adequately capture the contribution of electronic
excitation processes to the overall dynamics.

\section{\label{sec:level2}Computational Method}

The molecular dynamics in each set of simulations are modeled by real-time
time-dependent density-functional theory (TDDFT) on a real-space grid.  
The Kohn–Sham Hamiltonian has the following form:
\begin{equation}
\hat H_{\mathrm{KS}}(t)=
      -\frac{\hbar^{2}}{2m}\nabla^{2}
      +V_{\mathrm{ion}}(\mathbf r,t)
      +V_{H}[\rho](\mathbf r,t)
      +V_{XC}[\rho](\mathbf r,t),
\tag{1}
\end{equation}
where the first term, $-\hbar^{2}\nabla^{2}/2m$, is the single-electron kinetic-energy
operator.  
The electron density is  
\begin{equation}
\rho(\mathbf r,t)=
   \sum_{k=1}^{N_{\text{orbitals}}}
   2\bigl|\psi_{k}(\mathbf r,t)\bigr|^{2},
\tag{2}
\end{equation}
calculated by summing the contributions from all occupied Kohn–Sham orbitals.
$V_{\mathrm{ion}}$ denotes the external potential due to the ions, modeled with
norm-conserving pseudopotentials centered on each nucleus as given by Troullier
and Martins \cite{ref46}.
$V_{H}$ is the Hartree potential, representing the electrostatic Coulomb
interaction between the electrons,
\begin{equation}
V_{H}(\mathbf r,t)=
  \int \frac{\rho(\mathbf r^{\prime},t)}
            {|\mathbf r-\mathbf r^{\prime}|}\,
        d\mathbf r^{\prime},
\tag{3}
\end{equation}
The exchange-correlation potential V$_{XC}$ is 
approximated using the generalized gradient approximation (GGA), 
developed by Perdew et al. \cite{92PRB_GGA}.

Before the time-dependent calculations, we perform a
density-functional-theory (DFT) calculation to obtain the ground state of the
system, including the equilibrium electron density, self-consistent Kohn–Sham
orbitals, and the total energy.  With these initial conditions, we propagate the
orbitals in time using the time-dependent Kohn–Sham equation
\begin{equation}
i\hbar
\frac{\partial\psi_{k}(\mathbf r,t)}{\partial t}
      =\hat H_{\mathrm{KS}}(t)\,\psi_{k}(\mathbf r,t),
\tag{4}
\end{equation}
which is integrated with the propagator
\begin{equation}
\psi_{k}(\mathbf r,t+\delta t)=
   \exp\!\Bigl[-\,\tfrac{i}{\hbar}\hat H_{\mathrm{KS}}(t)\,\delta t\Bigr]
   \psi_{k}(\mathbf r,t)
\tag{5}
\end{equation}
and approximated by a fourth-order Taylor expansion:
\begin{equation}
\psi_{k}(\mathbf r,t+\delta t)\approx
   \sum_{n=0}^{4}\frac{1}{n!}
   \bigl(-\tfrac{i\delta t}{\hbar}\hat H_{\mathrm{KS}}(t)\bigr)^{n}
   \psi_{k}(\mathbf r,t).
\tag{6}
\end{equation}
The orbitals are propagated for $N$ time steps up to
$t_{\mathrm{final}} = N\Delta t$, where the time step $\Delta t$ is 
1~attosecond. This small timestep guarentees a conditionally stable time-propagation preserving the norm of the orbitals.
The $t_{\mathrm{final}}$ is set to 750 femtosecond, following Kwon \textit{et al.}\cite{ref16}

A uniform $60\times60\times60$ cubic grid with a spacing of
0.30~Å (an $18.0\times18.0\times18.0$~Å$^{3}$ box) represents the orbitals in
real space.  A 1~as time step and a 0.30~Å grid spacing have been shown to yield
accurate results in previous studies of Coulomb explosion.\cite{ref43,ref44}

The Kohn–Sham orbitals are set to zero at the box boundaries.  To prevent
unphysical reflections of electron density when fragments reach the edge, a complex absorbing
potential (CAP) surrounds the box.  Our simulations employ the form proposed by
Manopoulos\cite{ref48}:
\begin{equation}
-iw(x)= -\,\frac{i\hbar^{2}}{2m}
        \Bigl(\frac{2\pi}{\Delta x}\Bigr)^{2}
        f(y),\qquad
y=\frac{x-x_{1}}{\Delta x},
\tag{7}
\end{equation}
where $x_{1}$ and $x_{2}$ are the start and end of the absorbing region,
$\Delta x=x_{2}-x_{1}$, $c=2.62$ is a numerical constant, $m$ is the electron
mass, and
\begin{equation}
f(y)=\frac{4}{c^{2}}
      \left(\frac{1}{(1+y)^{2}}
            +\frac{1}{(1-y)^{2}}-2\right).
\tag{8}
\end{equation}

Ionic motion is treated classically within the Ehrenfest approximation.  The
electron-induced forces are
\begin{equation}
M_{i}\frac{d^{2}\mathbf R_{i}}{dt^{2}}=
  \sum_{j\neq i}^{N_{\text{ion}}}
      \frac{Z_{i}Z_{j}(\mathbf R_{i}-\mathbf R_{j})}
           {|\mathbf R_{i}-\mathbf R_{j}|^{3}}
  \;-\;
  \nabla_{\mathbf R_{i}}
  \int V_{\mathrm{ion}}(\mathbf r;\mathbf R_{i})\,
       \rho(\mathbf r,t)\,d\mathbf r,
\tag{9}
\end{equation}
where $M_{i}$, $Z_{i}$, and $\mathbf R_{i}$ are the mass, pseudocharge
(valence), and position of ion $i$, respectively, and
$N_{\text{ion}}$ is the total number of ions.  Equation~(9) is integrated with
the Verlet algorithm at every time step $\Delta t$.

Ground-state geometries were taken from the NIST~CCCBDB database and were used without further optimisation \cite{ref49,ref50}. The molecules were
placed at the centre of the simulation box to maximise the available simulation space.

To model double ionisation, we set the occupation of the highest occupied
molecular orbital (HOMO) to zero at $t=0$.  This mimics strong-field laser
double ionisation,\cite{ref8, ref9, ref10, ref12, ref16, ref19, ref20, ref21, ref22, ref25, ref28, ref30, ref34} which removes electrons almost instantaneously
(within one optical cycle)\cite{ref16}.  The resulting non-equilibrated hole evolves
together with the nuclei, so suddenly emptying the HOMO and propagating the
system in real time reproduces this non-adiabatic scenario.

Initial ionic velocities are sampled randomly from a Boltzmann distribution at
300~K.  This approach allows diverse fragmentation channels and provides a
holistic view of the fragmentation dynamics.  More than 190 simulations were
performed for each molecule, because formation of $\mathrm{H_{3}^{+}}$ is a
low-probability event among all dication dissociation pathways.

Our computational results fractional electron ejection can occur within the system, particularly during ionization processes. This behavior admits multiple interpretations. One perspective suggests that a specific fraction of electrons maintains localization within the molecular domain throughout the simulation timeframe, where the fractional charge may either undergo recombination with the ionized electron distribution or undergo dissociation over extended simulation periods. A more pragmatic interpretation views the non-integer charge as a statistical average: individual molecular fragments may preserve varying numbers of valence electrons, with some retaining one electron count while others maintain different values.

In the next section we present the ground-state HOMO obtained from DFT, typical
snapshots of the $\mathrm{H_{3}^{+}}$ formation mechanism, and histograms of the
branching ratios for the dissociation channels of ethane (C$_2$H$_6$), propane
(C$_3$H$_8$), and butane (C$_4$H$_{10}$).

\section{\label{sec:level3}Results and Discussion}

\subsection{\label{sec:level4}Ethane}

\begin{figure}[H]
  \centering
  \includegraphics[width=0.4\textwidth]{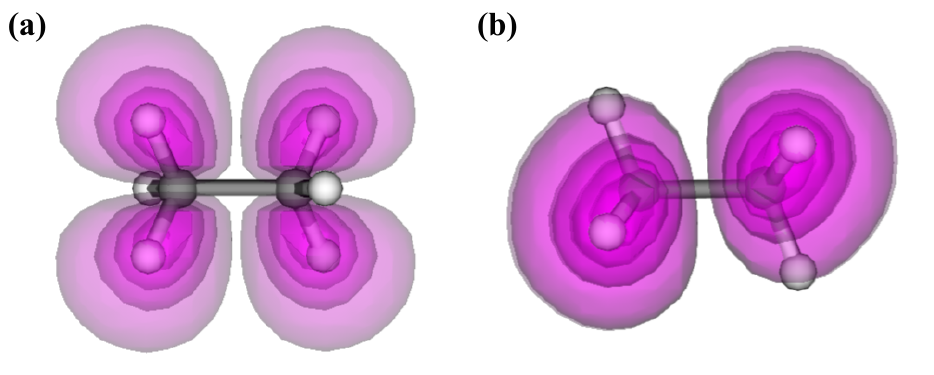}
  \caption{Four linearly spaced electron-density isosurfaces of the C$_2$H$_6$ HOMO. Views: (a) along the x-axis; (b) along the y-axis.}
  \label{fig1}
\end{figure}

Fig. \ref{fig1} shows that the HOMO of ethane is built from four
staggered $\sigma$(C-H) bond orbitals. Consequently, electron density
is concentrated on those four C-H bonds, while a nodal plane lies along the C-C axis, leaving virtually no density on the $\sigma$(C–C) bond. Therefore, removal of electrons from the HOMO selectively depletes electron density from the C–H bonds, weakening these bonds and facilitating H detachment.

\begin{figure}[H]
  \centering
\includegraphics[width=0.5\textwidth]{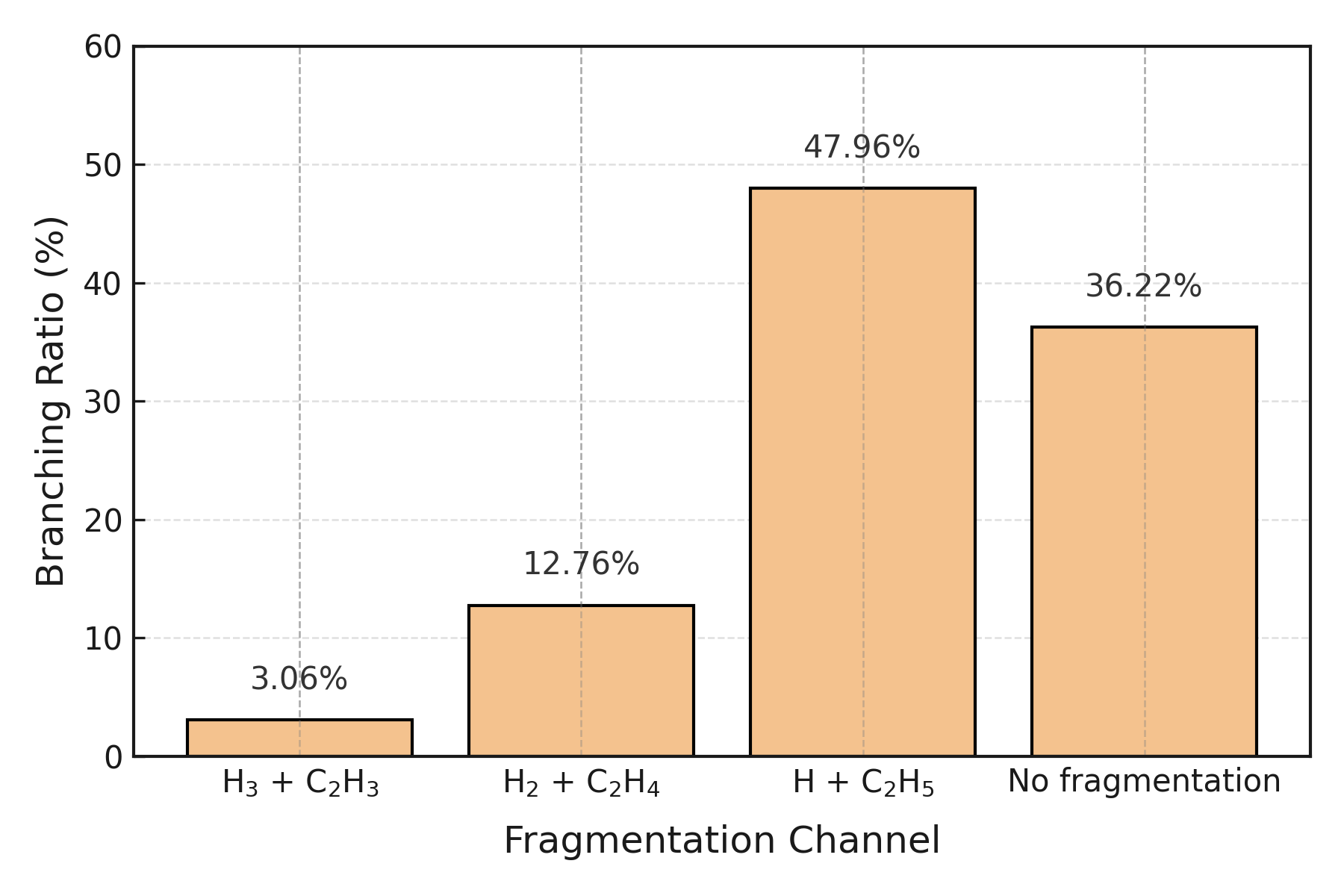}
  \caption{Fragmentation branching ratio of \ce{C2H6^2+} over the first 750fs, obtained by 196 TDDFT trajectories.}
  \label{fig2}
\end{figure}

Fig. \ref{fig2} shows the fragmentation branching ratios obtained from 196 TDDFT trajectories followed 
for the first 750 fs after double ionisation of \ce{C2H6}. No
trajectory exhibits C-C bond cleavage, in line with the negligible
$\sigma$(C-C) density predicted by the HOMO analysis. The dominant channel is
H + \ce{C2H5} (47.96\%); the departing hydrogen carries an average charge of +0.83 e, 
indicating that about 80 \% of the events produce \ce{H+} and 20 \% neutral H. 
The next most frequent pathway is \ce{H2 + C2H4} (12.76 \%). Here the \ce{H2} fragment 
bears an average charge of +0.39 e, implying an \ce{H2+} : \ce{H2} ratio of roughly 4 : 6. 
Formation of \ce{H3 + C2H3} is rare ($\approx$ 3\%); the \ce{H3} fragment carries a near-unit charge (+1.03 e).

Kraus \textit{et al.} did not quote a numerical \ce{H3+}/\ce{H+} ratio, but their time-of-flight spectra clearly show the \ce{H+} peak far larger than the \ce{H3+} peak,\cite{ref9} indicating proton loss dominates and \ce{H3+} is rare. Our TDDFT runs 
reproduce similar pattern: H + \ce{C2H5} is seen in about 50 \% of trajectories, while \ce{H3+} shows up only a few percent of the time.

Hoshina \textit{et al.} measured an \ce{H3+}/\ce{H2+} ratio of 0.94 using 50 fs, 795 nm pulses.\cite{ref8} Interpreting 
the fractional charges in our trajectories as population averages, we
obtain a ratio of 0.60. The difference is acceptable 
because the model starts from an instant HOMO vacancy, follows the motion for just 750 fs, and 
ignores higher-energy dication states. Despite those simplifications, the TDDFT approach still reproduces the 
main experimental trend obtained from strong-field laser and gets the
relative \ce{H3+} yield within the right order of magnitude.

\begin{figure}[H]
  \centering
  \includegraphics[width=0.5\textwidth]{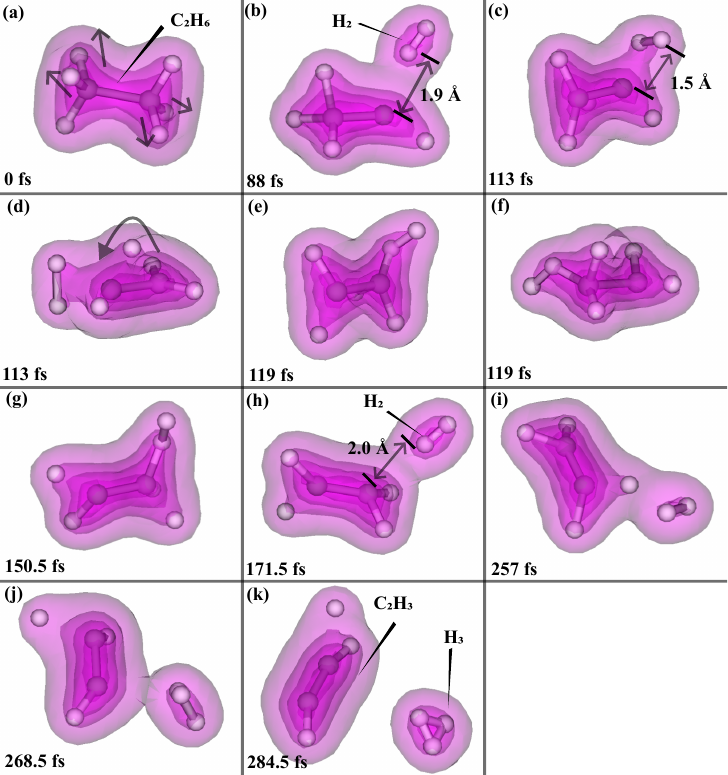}
  \caption{Snapshots showing a trajectory of the \ce{H3+} formation from \ce{C2H6^2+}. Snapshots (a)-(c), (e), (g)-(k) are viewed from the same angle. In the second row, (d) and (f) show the same times (113 fs and 119 fs) as (c) and (f), respectively, from a rotated viewpoint. Four linearly spaced electron-density isosurfaces are shown. \ce{H3+} is formed at 268.5 fs shown in (j).}
  \label{fig3}
\end{figure}

Fig. \ref{fig3} depicts a representative trajectory for \ce{H3+} formation from \ce{C2H6^2+}.  
After removal of the two HOMO electrons, all four corresponding C–H bonds elongate.  
Within \(\sim100\)\,fs an unstable quasi-neutral \ce{H2} fragment emerges; at 88\,fs (Fig. \ref{fig3}b)
the distance between this \ce{H2} and C2 is \(1.9\,\text{\AA}\)—significantly larger
than the \(\sim1.1\,\text{\AA}\) C–H bond length in a methyl group.  The H–H separation is
\(0.77\,\text{\AA}\), close to the \(0.74\,\text{\AA}\) of neutral \ce{H2}, and the fragment
bears a partial charge of \(+0.26\,e\).  Subsequent vibrations draw the \ce{H2} back towards
the carbon framework; by 113\,fs (Fig. \ref{fig3}c and \ref{fig3}d) the \ce{H2}–C2 distance contracts to
\(1.5\,\text{\AA}\) and one hydrogen migrates from C1 to C2.  At 119\,fs (Fig. \ref{fig3}e
and \ref{fig3}f) the migration is complete, giving the
\ce{[CH2-CH4]^2+} intermediate as proposed by
Kraus \textit{et al.}\cite{ref9} and Zhang \textit{et al.}\cite{ref11}
Over the next \(\sim50\)\,fs a more stable \ce{H2} fragment forms.  At 171.5\,fs (Fig. \ref{fig3}h)
it resides \(2.0\,\text{\AA}\) from the nearest hydrogen, carries \(+0.21\,e\), and retains an
H–H distance of \(0.77\,\text{\AA}\)—values closely matching the \(1.8\,\text{\AA}\), \(+0.20\,e\) and 0.77\,\text{\AA}\ calculated by Zhang \textit{et al.}\cite{ref11}
Another proton is drawn into proximity and then abstracted by the \ce{H2} (Fig. \ref{fig3}i),
reproducing the transition-state geometry suggested by Kraus \textit{et al.}\cite{ref9}
Finally, at 268.5\,fs \ce{H3+} (charge \(+1.02\,e\)) dissociates from the parent ion.(Fig. \ref{fig3}j)

\begin{figure}[H]
  \centering
  \includegraphics[width=0.5\textwidth]{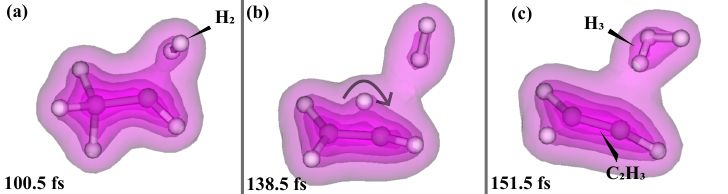}
  \caption{Snapshots showing an alternative trajectory of the \ce{H3+} formation from \ce{C2H6^2+}. Snapshots (a)-(c) are viewed from the same angle. Four linearly spaced electron-density isosurfaces are shown. \ce{H3+} is formed at 151.5 fs shown in (c).}
  \label{fig4}
\end{figure}

A single trajectory reveals a variant mechanism. Here, a stable \ce{H2} fragment detaches from C2 before any hydrogen migration occurs (Fig. \ref{fig4}a), whereas in the dominant pathway the nascent \ce{H2} is re-absorbed concurrently with the migration. The system therefore converts directly from the [\ce{H3C}–CH]$^2$$^+$···\ce{H2} arrangement to the [\ce{H2C}–\ce{CH2}]$^2$$^+$···\ce{H2} configuration through transfer of an H atom from C1 to C2 (Fig. \ref{fig4}b). Proton abstraction then follows rapidly, and \ce{H3+} is released at 151.5 fs (Fig. \ref{fig4}c)—earlier than in any other trajectory. 

All other trajectories leading to \ce{H3+} follow the first described
qualitative sequence, though formation times range from 190 fs to 680
fs owing to repeated H-scrambling and transient \ce{H2}
formation–reabsorption events before the key steps of H-migration,
\ce{H2} formation, and proton abstraction. Defining the formation time as the point at which the \ce{H3+} centre of mass begins to separate monotonically from that of the parent dication (similar to Zhang \textit{et al.}\cite{ref11}), we obtain an average of 354 $\pm$ 220 fs. The large standard deviation underscores the wide temporal window for this reaction and could also be attributed to the limited number of trajectories result in \ce{H3+} formation.

Our simulations reproduce the minimum-energy pathway (MEP) inferred from the stationary-point analyses of 
Kraus \textit{et al.}\cite{ref9} and Zhang \textit{et al}\cite{ref11}. Compared with the ab-initio molecular-dynamics 
(AIMD) study of Zhang \textit{et al.}, however, the present trajectories evolve on a noticeably 
longer timescale-a difference 
we attribute to the use of distinct electronic-structure methods
(adiabatic vs non-adiabatic). No evidence for \ce{H3+} formation via a 
roaming mechanism was found, possibly because of the limited number of trajectories or because 
the MEP dominates \ce{H3+} production from the ethane dication. Overall, these simulations offer 
a clear physical view of the \ce{H3+} formation from ethane.

\subsection{Propane}

\begin{figure}[H]
  \centering
  \includegraphics[width=0.4\textwidth]{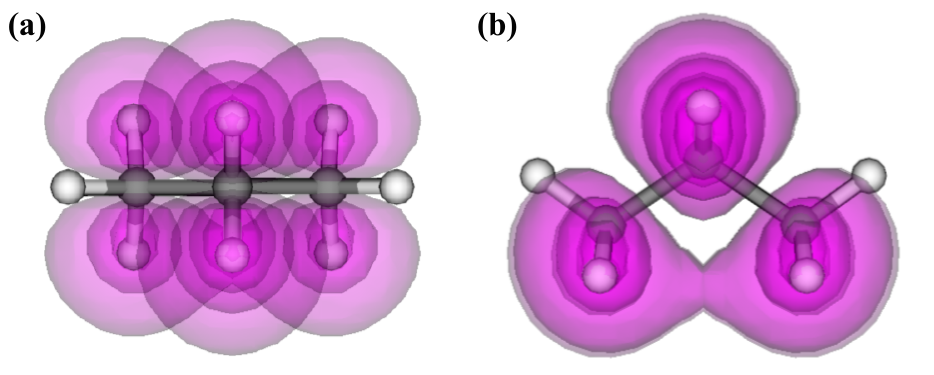}
  \caption{Four linearly spaced electron-density isosurfaces of the \ce{C3H8} HOMO. Views: (a) along the y-axis; (b) along the z-axis.}
  \label{fig5}
\end{figure}

Fig. \ref{fig5} shows that the HOMO of propane is constructed from six
$\sigma$(C-H) bond orbitals. Electron density is concentrated on all
six C-H bonds, with the four pseudo-axial (gauche-oriented) C-H bonds on the terminal 
carbons lying roughly perpendicular to the C-C skeleton, and with an even greater accumulation on the t
wo C-H bonds of the central \ce{CH2} group. A nodal plane coincides with the carbon chain, leaving 
virtually no density on either $\sigma$(C-C) bond. Consequently, ionisation from this HOMO selectively 
removes electron density from the C-H bonds—especially the central pair-weakening them and promoting 
hydrogen detachment while the C-C framework remains largely intact.

\begin{figure}[H]
  \centering
  \includegraphics[width=0.5\textwidth]{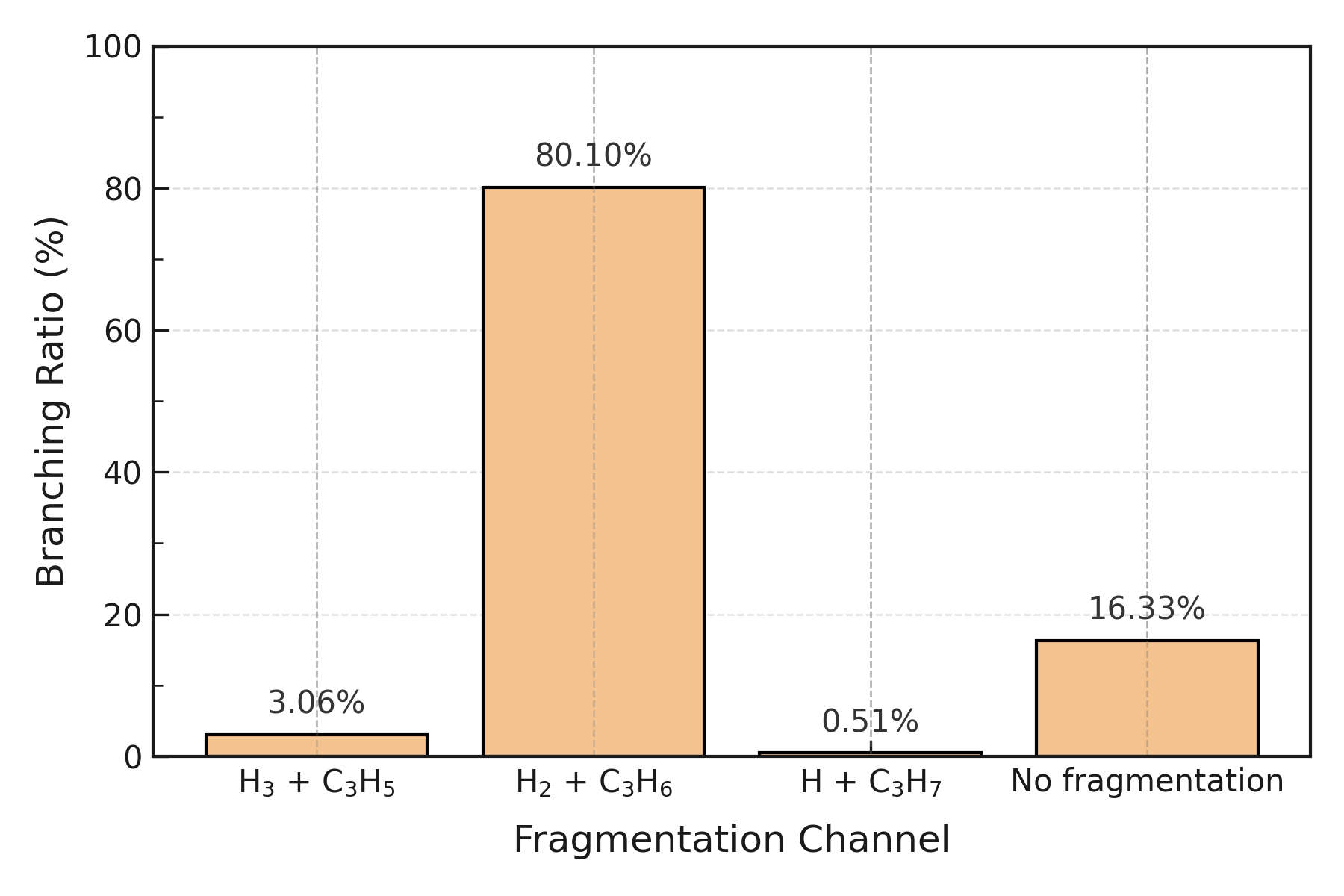}
  \caption{Fragmentation branching ratio of \ce{C3H8^2+} over the first 750fs, obtained by 196 TDDFT trajectories.}
  \label{fig6}
\end{figure}

Fig. \ref{fig6} summarizes the fragmentation branching ratios from 196 TDDFT 
trajectories of \ce{C3H8^{2+}} tracked for the first 750 fs after double ionisation.  
No C-C bond cleavage is ever observed, in agreement with the
negligible $\sigma$(C-C) density of the HOMO.  
The dominant channel is \ce{H2 + C3H6} (80.10 \%); the \ce{H2} fragment carries an average charge of \(+0.21\,e\), 
implying that roughly 20 \% of these events produce \ce{H2+} and 80 \% neutral \ce{H2}.  
Formation of \ce{H3 + C3H5} occurs in just 3.06 \% of trajectories, the same branching ratio as in the ethane dication; 
the \ce{H3} fragment bears \(+0.95\,e\), indicating it is almost equivalent to an \ce{H3+} ion.  
Direct \ce{H + C3H7} is very rare (0.51 \%), with the departing \ce{H} carrying \(+0.88\,e\). 
In 16.33 \% of simulations no fragmentation takes place within 750 fs.

\begin{figure}[H]
  \centering
  \includegraphics[width=0.5\textwidth]{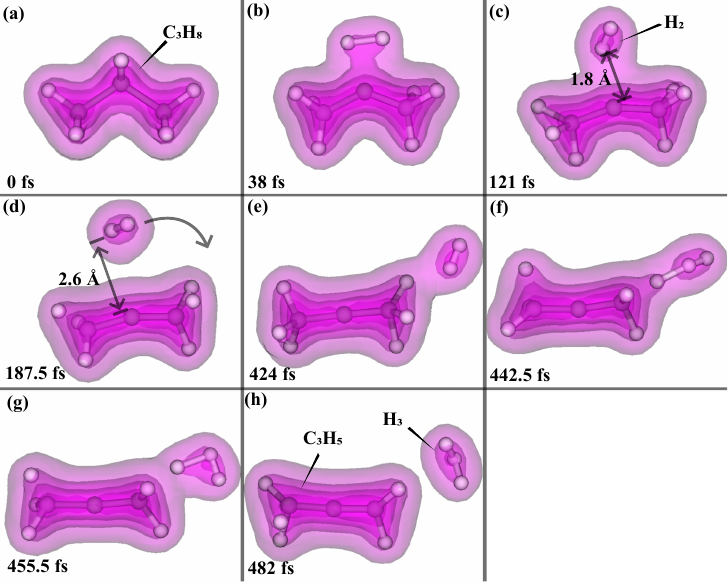}
  \caption{Snapshots showing a trajectory of the \ce{H3+} formation from \ce{C3H8^2+}. Snapshots (a)-(h) are viewed from the same angle. Four linearly spaced electron-density isosurfaces are shown. \ce{H3+} is formed at 455.5 fs shown in (g).}
  \label{fig7}
\end{figure}

Fig. \ref{fig7} illustrates a representative \ce{H3+}-formation trajectory from \ce{C3H8^{2+}}.  
Removal of the two HOMO electrons induces elongation of the six
constituent $\sigma$(C-H) bonds, with the pair on the central C2 stretching most dramatically. 
During the first 100\,fs these two hydrogens repeatedly form transient \ce{H2} moieties, 
as shown in Fig. \ref{fig7}b, that promptly re-absorb—an oscillatory behaviour akin to that 
seen in \ce{C2H6^{2+}}. At $\approx120$\,fs (Fig. \ref{fig7}c) a stable \ce{H2} fragment 
emerges 1.8\,\AA\ from C2, with an H-H distance of 0.71\,\AA\ (reflecting vibrational compression) 
and a \(+0.25\,e\) charge; this event signals the onset of \ce{H2} roaming. Between Fig. \ref{fig7}b 
and \ref{fig7}c, during the formation of this stable quasi-neutral
\ce{H2} fragment, the C1-C2-C3 angle opens markedly and the two \ce{C-C} bonds become nearly collinear, r
eflecting partial \textit{sp} re-hybridisation at C2; this electronic/geometric relaxation accompanies, 
and likely facilitates, the subsequent \ce{H2} roaming. Over the next few tens of femtoseconds, 
the \ce{H2} migrates away from its formation site; by 187.5\,fs (Fig. \ref{fig7}d) it resides 
2.6\,\AA\ from the nearest carbon, bears \(+0.17\,e\), and has an H-H length of 0.78\,\AA. 
Thereafter, across the ensuing \(\sim250\)\,fs, the roaming \ce{H2} approaches the C3 site, 
abstracts its adjacent proton, and at 455.5\,fs generates \ce{H3+} (Fig. \ref{fig7}e–g). 
The newly formed \ce{H3+}, carrying \(+1.03\,e\), then dissociates from the parent dication.

All \ce{H3+}‐forming trajectories adhere to this sequence, although the overall formation time 
spans 210–570 fs. This variability stems from differences in the durations of transient \ce{H2} 
formation–reabsorption (50–450 fs) and \ce{H2} roaming (80–260 fs) before the final steps of 
stable \ce{H2} formation, roaming to a terminal carbon, and proton abstraction. Defining the 
formation time exactly as for ethane, i.e., the instant when the \ce{H3+} center of mass begins 
its monotonic separation from the dication-yields an average formation
time of 402 $\pm$ 141 fs.

\subsection{Butane}

\begin{figure}[H]
  \centering
  \includegraphics[width=0.4\textwidth]{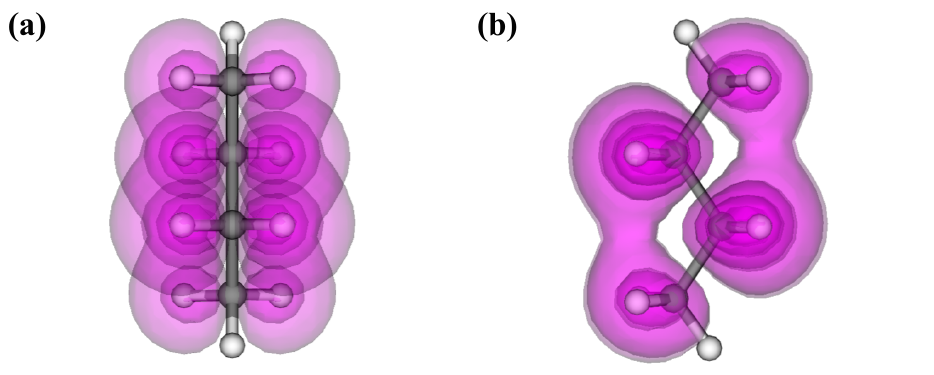}
  \caption{Four linearly spaced electron-density isosurfaces of the \ce{C4H10} HOMO. Views: (a) along the x-axis; (b) along the z-axis.}
  \label{fig8}
\end{figure}

As Fig. \ref{fig8} shows the HOMO of butane is constructed from eight
$\sigma$(C-H) bond orbitals. The electron density is concentrated on all
eight C-H bonds: the four staggered (gauche-oriented) C–H bonds on the terminal
methyl groups C1 and C4, and the four C-H bonds on the internal methylene carbons 
C2 and C3, the latter displaying the highest amplitude. A nodal plane coincides with 
the C-C backbone, leaving virtually no density on either $\sigma$ (C-C) bond. Minor lobes of density extend into the interstitial regions between the C1–H/C3–H and C2–H/C4–H bond pairs, indicating weak through-space interaction among remote $\sigma$ (C–H) orbitals.

\begin{figure}[H]
  \centering
  \includegraphics[width=0.5\textwidth]{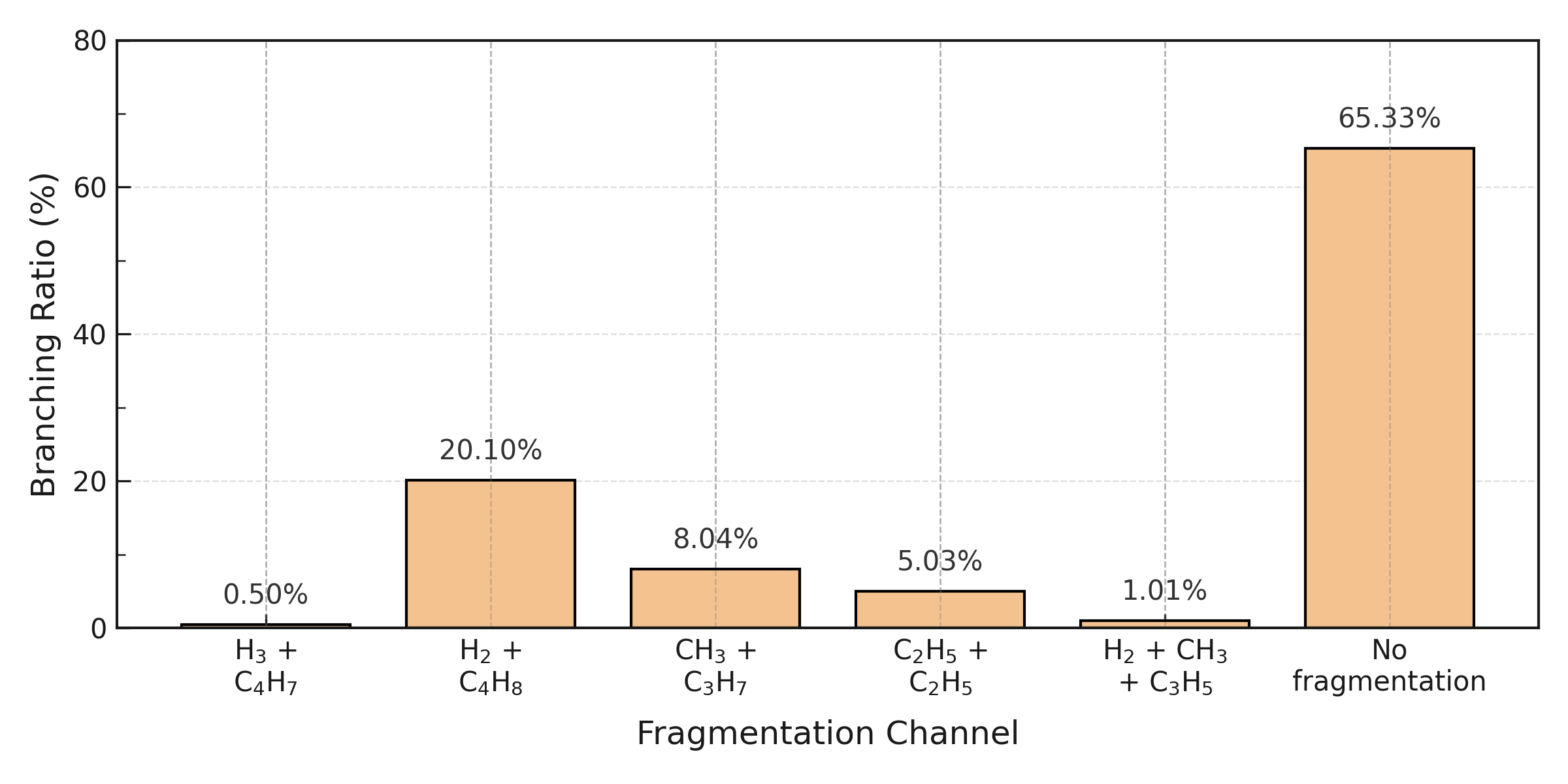}
  \caption{Fragmentation branching ratio of \ce{C4H10^2+} over the first 750fs, obtained by 200 TDDFT trajectories.}
  \label{fig9}
\end{figure}

Fig. \ref{fig9} illustrates the fragmentation branching ratios from 200 TDDFT trajectories of \ce{C4H10^2+} followed for the first 750 fs after double ionisation. The most probable outcome is no fragmentation within the simulated window (65.33 \%) which reflects reflects charge delocalisation over the longer C4 chain. Among reactive events, the chief pathway is \ce{H2 + C4H8} (20.10 \%). The \ce{H2} fragment carries an average charge of +0.17 e, 
implying that $\approx$ 20 \% of these events yield \ce{H2+} and
$\approx$ 80 \% neutral \ce{H2}. A very small fraction of trajectories (0.50 \%) form \ce{H3 + C4H7}; 
the nascent \ce{H3} bears an almost integer charge (+0.989 e). All other observed channels—\ce{CH3 + C3H7}, 
\ce{C2H5 + C2H5}, and \ce{H2 + CH3 + C3H5}—collectively account for 14.1 \% of the statistics and each requires 
at least one C-C bond cleavage. Although carbon–carbon fission is not the theoretical focus of the present work, 
its appearance here (in contrast to ethane and propane) is noteworthy: the longer carbon chain provides more accessible $\sigma$(C–C) antibonding combinations once two electrons are removed, thereby lowering the barrier to backbone rupture.

\begin{figure}[H]
  \centering
  \includegraphics[width=0.5\textwidth]{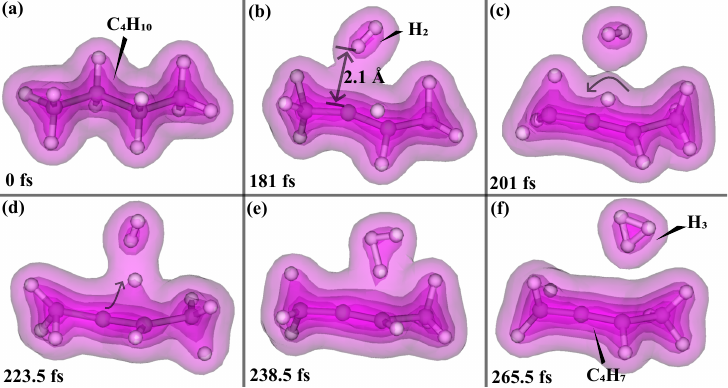}
  \caption{Snapshots showing a trajectory of the \ce{H3+} formation from \ce{C4H10^2+}. Snapshots (a)-(f) are viewed from the same angle. Four linearly spaced electron-density isosurfaces are shown. \ce{H3+} is formed at 238.5 fs shown in (e).}
  \label{fig10}
\end{figure}

Fig. \ref{fig10} presents the only trajectory in which \ce{C4H10^2+} forms \ce{H3+}. Removal of the two HOMO electrons lengthens all eight constituent $\sigma$(C–H) bonds, the four on the internal carbons (C2, C3) stretching most strongly. For the first 170 fs, the system displays only bond vibrations; transient \ce{H2} formation is absent, consistent with greater charge delocalisation along the C4 chain. At 181 fs (Fig. \ref{fig10}b) a stable \ce{H2} fragment appears 2.1 Å from C2, with an H–H distance of 0.80 Å and a +0.13 e charge—marking the onset of the reaction. The quasi-neutral \ce{H2} remains near its formation site for $\approx$40 fs. At 201 fs (Fig. \ref{fig10}c) a hydrogen migrates from C3 to C2; over the next ~20 fs this proton is drawn toward the \ce{H2} (Fig. \ref{fig10}d). \ce{H3+}, carrying +0.99 e, is fully formed and detaches at 238.5 fs (Fig. \ref{fig10}e). During this interval the C1–C2–C3 angle at the ejection site widens toward linearity, mirroring the backbone straightening observed in propane and reflecting partial sp-like rehybridisation of the central carbon.

\begin{figure}[H]
  \centering
  \includegraphics[width=0.5\textwidth]{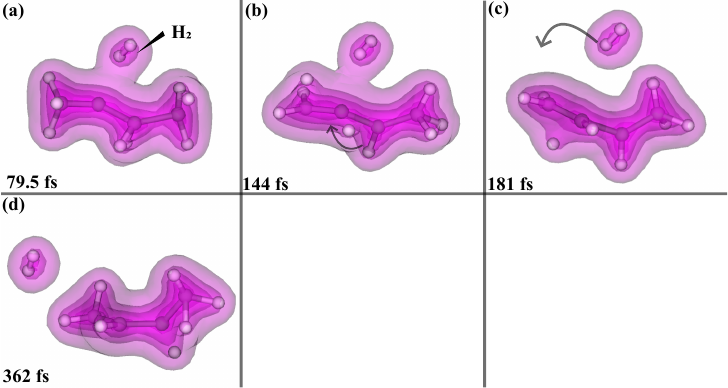}
  \caption{Snapshots showing a trajectory of the \ce{H2} roaming from \ce{C4H10^2+}. Snapshots (a)-(d) are viewed from the same angle. The simulation results in \ce{H2} ejection. Four linearly spaced electron-density isosurfaces are shown. }
  \label{fig11}
\end{figure}

The \ce{H2} fragment described above survives for only 50 fs and never strays far from its formation site; it therefore does not meet the criteria for \ce{H2} roaming. True roaming is nonetheless observed in a few simulations. In the trajectory depicted in Fig. \ref{fig11}, a stable  \ce{H2} fragment is expelled from C2 at 79.5 fs (Fig. \ref{fig11}a) and remains nearby until 144 fs, when a hydrogen migrates from C3 to C2 (Fig. \ref{fig11}b). Over the next $\approx$ 220 fs the  \ce{H2} wanders extensively around the molecular framework. By 362 fs it approaches the hydrogen on C1 but fails to abstract a proton; the outcome is the ejection of an  \ce{H2} rather than the formation of \ce{H3+}.

\subsection{Comparison Between the Three Molecules}

Across the 196-trajectory ensembles for both ethane and propane, the branching ratio for \ce{H3+} formation is
\(6/196 = 3.1\%\), with a Wilson 95\,\% confidence interval (CI) of \(1.4\text{–}6.5~\%\).
In butane, only one productive event occurs in 200 trajectories, yielding \(0.5~\%\)
(95\,\% CI: \(0.09\text{–}2.8~\%\)); its smaller value and wider interval underscore the much lower—and less certain—
probability of \ce{H3+} formation.  Mean formation times reveal no simple trend:
\ce{H3+} appears, on average, sooner in ethane than in propane, but both means carry large standard deviations,
and the lone butane event is shorter than either mean.

Ethane reacts chiefly along the minimum-energy path (MEP):
an initial hydrogen migration forms the [\ce{CH2-CH4}]$^2$$^+$ intermediate,
a quasi-neutral \ce{H2} fragment departs from the \ce{CH4} side,
and then abstracts a nearby proton to give \ce{H3+}.
A minor channel in which a stable \ce{H2} forms first—and then captures the proton supplied by subsequent
H migration—was also observed.
No roaming-\ce{H2} mechanism surfaced for ethane.

Propane is dominated by the \ce{H2}-roaming route:
a quasi-neutral \ce{H2} fragment forms at C2, roams toward either terminal carbon,
and abstracts a proton to produce \ce{H3+}.
In butane, the single productive trajectory mirrors ethane’s minor pathway:
\ce{H2} forms at C2 and reacts with a proton that has migrated from C3.
Although roaming \ce{H2} occurs in butane, it did not lead to \ce{H3+} within the present ensemble,
so its role cannot yet be excluded.

The virtually identical yields of ethane and propane, followed by the sharp drop in butane,
contrast with the monotonic decline reported for the alcohol series \cite{ref20}.
Propane’s relatively high yield likely arises from the facile formation of quasi-neutral \ce{H2} at the
central carbon, for which two equivalent pathways for proton abstraction exist.
In butane, charge delocalisation on longer carbon chain and competing
low-energy \ce{C-C} bond-cleavage channels appear to suppress
\ce{H3+} formation.

\section{Conclusion}

This study delivers a real-time, time-dependent density-functional-theory (TDDFT) picture of \ce{H3+} formation in alkane dications. By 
suddenly emptying the HOMO and propagating electrons and nuclei together on attosecond time steps, the simulations capture 
non-adiabatic charge migration and nuclear motion without relying on pre-computed potential-energy surfaces, as in previous BOMD studies.

For ethane, our TDDFT simulations qualitatively capture the experimental fragmentation pattern. The \ce{H+}+\ce{C2H5^{2+}} channel 
dominates over the \ce{H3+}+\ce{C2H3^{2+}} channel \cite{ref9}, and the simulated \ce{H3+}/\ce{H2+} yield ratio of 0.60 is reasonably close to 
the experimental value of 0.94 \cite{ref8}. The simulations give an overall branching ratio of $\approx$ 3\% and the reaction proceeds 
along the minimum-energy pathway  identified in earlier
stationary-point studies \cite{ref9,ref11}. One trajectory also exposes a shortened variant that bypasses 
the [\ce{CH2}-\ce{CH4}]$^2$$^+$ intermediate.
Propane forms \ce{H3+} almost exclusively through \ce{H2} roaming: a quasi-neutral \ce{H2} formed at C2 
wanders to either terminal carbon before abstracting a proton.
Despite their mechanistic contrast, ethane and propane yield indistinguishable \ce{H3+} branching 
ratios \((6/196 \approx 3.1~\%;\ \text{Wilson 95\% CI}=1.4\text{–}6.5~\%)\),
whereas butane falls to \(0.5~\%\) \((1/200;\ \text{CI}=0.09\text{–}2.8~\%)\).
The sharp drop originates from low-barrier \ce{C-C} cleavage channels in the larger dication, 
while the single \ce{H3+} formed arises when an \ce{H2} fragment generated at C2 abstracts a proton that 
has migrated from C3. In our ethane and butane simulations we did not observe \ce{H3+} formation via the 
\ce{H2}‐roaming pathway, most likely because the present TDDFT implementation cannot capture proton 
tunneling—a quantum effect that would otherwise raise the probability of \ce{H3+} production along this route. 

These findings show that \ce{H3+} yield in alkanes is governed not by chain length alone but by the competition between  
(i) ease of ejecting a quasi-neutral \ce{H2}
(ii) the extent of charge delocalisation on the carbon chain and
(iii) the number of alternative fragmentation routes opened by double ionisation.  
TDDFT thus emerges as a powerful complement to experiment for mapping ultrafast ion chemistry.

Future work should therefore expand the trajectory set, include explicit laser fields, and combine higher-level electronic-structure methods with targeted experiments to validate the mechanistic crossover between MEP-dominated (ethane) and \ce{H2}-roaming-dominated (propane) regimes, and to clarify whether longer alkanes can access \ce{H2}-roaming-mediated \ce{H3+} formation at all.

\section{acknowledgments}

\begin{acknowledgments} 
This work was supported by the Natural Science Foundation
of Henan Province under Grant No. 252300421490, and by the National
Science Foundation (NSF) 
under Grant No. DMR-2217759. Computational resources were provided by
ACES at 
Texas A$\&$M University through allocation PHYS240167 from the 
Advanced Cyberinfrastructure Coordination Ecosystem: Services $\&$
Support (ACCESS) program, 
supported by NSF grants 2138259, 2138286, 2138307, 2137603, and 2138296.
\end{acknowledgments}

\section*{Data Availability Statement}
The data that support the findings of this study are available
from the corresponding author upon reasonable request.

\section*{AUTHOR DECLARATIONS}
\par\noindent
{\bf Conflict of Interest}

The authors have no conflict of interest to disclose.

%

\end{document}